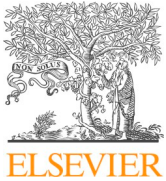
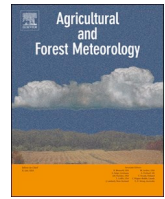

# Influence of climate variability on the potential forage production of a mown permanent grassland in the French Massif Central

Iñigo Gómara[a,*], Gianni Bellocchi[b], Raphaël Martin[b], Belén Rodríguez-Fonseca[c], Margarita Ruiz-Ramos[a]

[a] CEIGRAM/Departamento de Producción Agraria, Universidad Politécnica de Madrid, Paseo Senda del Rey 13, 28040 Madrid, Spain
[b] UCA, INRA, VetAgro Sup, Unité Mixte de Recherche sur Écosystème Prairial (UREP), Site de Crouel 5, Chemin de Beaulieu, 63000 Clermont-Ferrand, France
[c] Departamento de Física de la Tierra y Astrofísica, Universidad Complutense de Madrid, Plaza de Ciencias 1, 28040 Madrid, Spain

ABSTRACT

Climate Services (CS) provide support to decision makers across socio-economic sectors. In the agricultural sector, one of the most important CS applications is to provide timely and accurate yield forecasts based on climate prediction.

In this study, the Pasture Simulation model (PaSim) was used to simulate, for the period 1959–2015, the forage production of a mown grassland system (Laqueuille, Massif Central of France) under different management conditions, with meteorological inputs extracted from the SAFRAN atmospheric database. The aim was to generate purely climate-dependent timeseries of optimal forage production, a variable that was maximized by brighter and warmer weather conditions at the grassland.

A long-term increase was observed in simulated forage yield, with the 1995–2015 average being 29% higher than the 1959–1979 average. Such increase seems consistent with observed rising trends in temperature and $CO_2$, and multi-decadal changes in incident solar radiation. At interannual timescales, sea surface temperature anomalies of the Mediterranean (MED), Tropical North Atlantic (TNA), equatorial Pacific (El Niño Southern Oscillation) and the North Atlantic Oscillation (NAO) index were found robustly correlated with annual forage yield values. Relying only on climatic predictors, we developed a stepwise statistical multi-regression model with leave-one-out cross-validation. Under specific management conditions (e.g., three annual cuts) and from one to five months in advance, the generated model successfully provided a *p*-value < 0.01 in correlation (*t*-test), a root mean square error percentage (%RMSE) of 14.6% and a 71.43% hit rate predicting above/below average years in terms of forage yield collection.

This is the first modeling study on the possible role of large-scale oceanic–atmospheric teleconnections in driving forage production in Europe. As such, it provides a useful springboard to implement a grassland seasonal forecasting system in this continent.

## 1. Introduction

Nowadays, a great demand exists for translating the large amount of climate data and information into customized tools, products and services (European Commission, 2014). These tools, so-called Climate Services (CS), provide support to decision makers across socio-economic sectors (e.g., water management, agriculture, energy production). In the agricultural sector, one of the most important CS applications is to provide timely and accurate yield forecasts based on climate prediction, which can be used to adopt best management practices and readily incorporate them into on-farm management plans in a cost-effective manner (Lemos et al., 2012; Mase and Prokopy, 2014). User-driven climate-information services may also serve to prevent emergencies in target food-insecure regions (e.g. Haile, 2005) or orient market forces and regulations related to food production (e.g. Whitfield et al., 2018), and support environmental protection while adapting to climate change (e.g. Hackenbruch et al., 2017).

Accurate weather predictions beyond a two-week horizon are not possible because of the intrinsic chaotic behavior of the Earth's atmospheric system (Lorenz, 1963; 1982; Yoden, 2007). However, seasonal predictions can be retrieved using as input oscillatory and slowly varying components in the coupled ocean-atmosphere-land system that dominate the weather response (Luo and Wood, 2006). Sea surface temperatures (SSTs) are likely the most important source of climate predictability at multiple timescales (e.g., seasonal, interannual, multi-decadal) because ocean water has a much higher heat capacity than air (Bjerknes, 1969). SST anomalies over large marine areas can affect the atmospheric circulation, so that some regions may become warmer and wetter than normal, while others become cooler and drier during specific periods (Hurrell, 2008). SST anomalies thus provide valuable information on the evolution of air-sea interactions and associated impacts in climate, in both nearby and distant locations (Hand et al., 2018), as these anomalies can alter and propagate into oceanic-atmospheric pathways (e.g. Wallace and Gutzler, 1981; Årthun et al., 2017; Gleixner et al., 2017; Mamalakis et al., 2018). This propagation takes place through atmospheric large-scale recurrent and persistent anomaly patterns, which are called teleconnection patterns (hereafter referred as "teleconnections" for simplicity; Barnston and Livezey, 1987).

In the Euro-Atlantic sector, the North Atlantic Oscillation (NAO) is the most prominent teleconnection affecting climate variability on a broad frequency band (daily to multi-decadal). The NAO characterizes the mean sea level pressure (mslp) gradient between sub-polar and sub-tropical latitudes and is driven by multiple variability sources (like stochastic internal, oceanic, solar, stratospheric; Hurrell et al., 2003; Pinto and Raible, 2012). Under its positive phase (NAO+), synoptic weather systems are deflected northward, causing cloudier, wetter conditions over northern Europe and sunnier, drier conditions over southern Europe (Gómara et al., 2014). The reverse holds for NAO-.

The warming or cooling of Pacific Equatorial SSTs, known as El Niño Southern Oscillation (ENSO), is the major source of climate variability affecting different regions of the world (e.g. Trenberth, 1997). At interannual timescales, a positive ENSO phase tends to be associated with NAO- in late winter (Brönnimann, 2007), and with increased precipitation over southern Europe during summer and fall (Shaman, 2014), although this link appears to be non-stationary in time (López-Parages and Rodríguez-Fonseca, 2012). Additional SST patterns with influence over Europe are the Tropical North Atlantic (TNA) and the Mediterranean (MED). A warmer TNA is known to increase precipitation over southwestern Europe in winter (Okumura et al., 2001; Czaja and Frankignoul, 2002; Rodríguez-Fonseca et al., 2006; Losada et al., 2007). A warmer MED is linked to higher temperatures over Europe in summer and autumn (Feudale and Shukla, 2007; García-Serrano et al., 2013). At lower frequencies, the Atlantic Multidecadal Oscillation (AMO; a multidecadal warming/cooling of North Atlantic SSTs) is known to influence the NAO (Gastineau and Frankignoul, 2015), the north Atlantic storm track (Woollings et al., 2012; Gómara et al., 2016; Diodato and Bellocchi, 2017; Vaideanu et al., 2017) and Euro-Mediterranean precipitation (López-Parages and Rodríguez-Fonseca, 2012; Diodato and Bellocchi, 2018). Thus, teleconnections can provide prediction of key variables for agro-ecosystems like maximum/minimum daily temperatures, incident solar radiation, precipitation or wind speed, which can be assimilated into monthly-to-decadal forecasting systems (e.g. Dannenberg et al., 2018).

Decades of research have disclosed strong connections between ENSO and major crop yields worldwide. Cane et al. (1994) found a strong anticorrelation between ENSO index and maize yields in Zimbabwe (even stronger than for rainfall). Phillips et al. (1999) found a positive effect of El Niño episodes in the U.S. corn belt, a result subsequently corroborated by Kellner and Nigoyi (2015). Meinke and Hochman (2000) showed that positive ENSO events depressed rainfall in Eastern Australia, leading to lower-than-average wheat yields. Other SST indices were investigated for links to agricultural yield: for example, de la Casa and Ovando (2014) showed a significant correlation between the AMO index and corn and soybean in Argentina. To date, most investigations of the effect of large-scale oceanic/atmospheric variables on agriculture in Europe have been restricted to ENSO (e.g. Cantelaube et al., 2004; Capa-Morocho et al., 2014, 2016a,b; Ceglar et al., 2017) and TNA phenomena (Capa-Morocho et al., 2016b). Considering SSTs and additional observed weather inputs, the Crop Growth and Monitoring System, developed by the European Commission Joint Research Center within the Monitoring Agricultural Resources (JCR-MARS) activities, provides short-term (in-season) yield forecasts of the main food crops in Europe (Vossen and Rijks, 1995; Lazar and Genovese, 2004; de Wit et al., 2005; Pagani et al., 2017). For grasslands, biomass production is monitored from remote sensing products, but Europe-wide forecasts are not provided yet (MARS, 2019).

Permanent grasslands are agricultural lands used for cultivating "grasses or other herbaceous forage", through either self-seeding or sowing operations, not included in the crop rotation of the farm in at least five years [Reg. EU 2017/2393]. These ecosystems hold a remarkable part of Europe's biodiversity and provide a wide variety of natural resources for human consumption (e.g. livestock, meat and dairy products). Around 20% of the total EU-28 territory is covered by grasslands. Grasslands provide 25% of total food intake for European livestock and play a key role in the greenhouse gas (GHG) budget (Hörtnagl et al., 2018). In France, permanent grasslands cover 36% of agricultural land area and this number can be as high as ∼60% in some regions like the Massif Central (Graux et al., 2013). This region, which has the highest number of dairy farms in the country, significantly contributes to keep France ranked amongst the main EU-28 producers of bovine meat and milk (Eurostat, 2019). Considering the importance of the livestock sector in the region, and its dependency on forage resources for animals, there is a great need to develop forage prediction tools to assist farmers in managing risk for grassland-livestock production systems. Establishing a framework for yield prediction matching grassland management, weather and climate is particularly relevant in areas like the Massif Central of France (focus of this study), whose economy is vulnerable to climate variability and change (e.g. Virto et al., 2015; Hamidov et al., 2018).

The impact of climate on grassland productivity has been evaluated through field campaigns with *in-situ* meteorological measurements (Black et al., 2006; Dürr et al., 2015; Chen et al., 2017) and remote sensing products (Kawabata et al., 2001; Ding et al., 2017; Liu et al., 2017). Overall in the mid-latitudes, mild temperatures and generous precipitation are linked to increased biomass production (Menzi et al., 1991). Nevertheless, these factors can notably vary depending on local climatic conditions (e.g. semi-arid, humid, mountainous; Nippert et al., 2006; Yang et al., 2008; Carlyle et al., 2014; Jones et al., 2016). Most of these studies have provided high-quality grassland data but were limited by the short periods of time over which the analyses were carried out. A remarkable exception can be found in Chen et al. (2017), which presented forage yield data over 1939–2016. This is an issue because extended timeseries are scarce but essential to identify robust statistical links between climate variability and yield responses in any agricultural system (Craine et al., 2012; Capa-Morocho et al., 2014; 2016a,b). As an alternative, crop simulation models can be used to generate long-term timeseries of grassland performance data.

Grassland primary production and GHG budgets have been quantified in modeling studies based on observational flux data at specific locations in Europe (Ma et al., 2015), and world-wide (Ehrhardt et al., 2018). Such studies have substantiated the use of grassland biogeochemical models at local, regional and continental scales (Vuichard et al., 2007a) and shown their potential to represent grassland systems under a variety of conditions (Brilli et al., 2017).

Additional opportunities are nowadays offered by high-resolution long-term datasets of observational atmospheric records (e.g., ERA-WATCH, SAFRAN; Weedon et al., 2010; Vidal et al., 2010), which provide support to model intercomparison initiatives (e.g., MACSUR, AgMIP, CN-MIP or ISIMIP; Frieler et al., 2017). It is now possible to feed grassland-specific numerical models with long-term weather observations and perform simulations under potential conditions, i.e. optimal management (N fertilization, number of cuts) and biotic stresses effectively controlled (no pests and diseases). As a result, long-term purely climate-dependent timeseries of potential forage yield can be generated and analyzed with grassland models. This can be done at the cost of downgrading the quality of outputs (i.e. data can be coarse grained for simulation purposes, e.g. Hoffmann et al., 2016). The issue here is to ensure that this downgrading is still adequate for delivering operationally relevant CS.

Based on this novel modeling methodology, a framework was developed in this study to support CS applications and tools with a focus on grassland ecosystems. In particular, we explored the potential links between climate variability, teleconnection patterns and harvested forage in a permanent grassland located in the Massif Central of France. The aim was to provide evidence that skillful seasonal forecasts of forage productivity could be attainable using only oceanic and atmospheric predictors.

## 2. Data

### 2.1. Grassland site

The permanent grassland system in this study is located in Laqueuille (45° 38′ N, 2° 44′ E, 1040 m a.s.l.), in a west-facing leeward slope of the French Massif Central (Fig. 1a and S1a). The grassland plant community is a grass-clover mix dominated by perennial ryegrass (*Lolium perenne* L.) and white clover (*Trifolium repens* L.). The soil is a 1 m deep Andosol with 11% carbon and 18% organic matter in the 0.1 m topsoil (they rapidly decline further down) and silt loam texture (26% sand, 20% clay). Soil field capacity, permanent wilting point and saturation are, respectively, 0.36, 0.22 and 0.53 $m^3 \ m^{-3}$ (Klumpp et al., 2011). The grassland is a long-term observation system for research and experimentation managed by INRA since 2003 (SOERE, 2019).

### 2.2. Gridded oceanic and atmospheric databases

Gridded oceanic and atmospheric databases were used to analyze climate variability at global, European and country (France) scales. All the information regarding these datasets is summarized in Table S1.

### 2.3. Meteorological observations at the site

An INRA meteorological station at the grassland site provides hourly values of surface air temperature, relative humidity, precipitation, wind speed and incident solar radiation. Due to its mountainous location, the grassland site presents sub-arctic climatic conditions (Dfc Köppen-Geiger classification), with limited sunlight, cold temperatures and abundant precipitation year-round (cf. Table 1 and Fig. 1b). The station records cover the period 1996–2015, but the extensive data gaps in the record of solar radiation only made it possible to perform grassland simulation over eight years (2008–2015). Instead, we referred to the SAFRAN atmospheric re-analysis (cf. Table S1) to obtain long-term meteorological observations of the area. To this aim, the timeseries of the closest SAFRAN grid cell (central point) to the simulated grassland were selected (3.64 km away, similar altitude). To allow comparison between *in-situ* and SAFRAN measurements, the maximum overlapping period of both datasets was selected for all weather variables (1996–2015), except for surface incident solar radiation (2008–2015). Table 1 provides the mean and standard deviation (SD) values of both datasets.

On average, a satisfactory agreement was observed between station and SAFRAN data (Table 1). Despite its generally good agreement with Laqueuille station data and its validation throughout France (Quintana-Seguí et al., 2008), SAFRAN has several systematic biases that deserve further consideration: it slightly overestimates srad, Pr, and T, while slightly underestimating wind speed (wspd). According to Fig. 1b, biases in Pr and wspd are larger during the winter months, and it can be logically assumed that their potential impacts on soil water balance, evapotranspiration and simulated forage yield should be limited, considering that grassland biomass production is minimal from November to March due to low temperatures. For srad, the largest biases arise during spring and summer. These differences could be associated with upslope or local radiation fogs at the grassland site (with plants acting as an important source of water vapor), not adequately captured by SAFRAN. For standard deviation, lower values are obtained in SAFRAN for T, relative humidity and wspd at daily and interannual timescales compared to the INRA station (Table 1). Contrastingly, srad variance is overestimated in SAFRAN. Indeed, in the computation of anomalies for this study, all means were removed, thus minimizing the potential influence of biased data in the final results. As a consequence, only the SAFRAN weather data (57 years) were utilized as an input for grassland modeling in this study, discarding the weather station data owing to their short temporal extent (8 years).

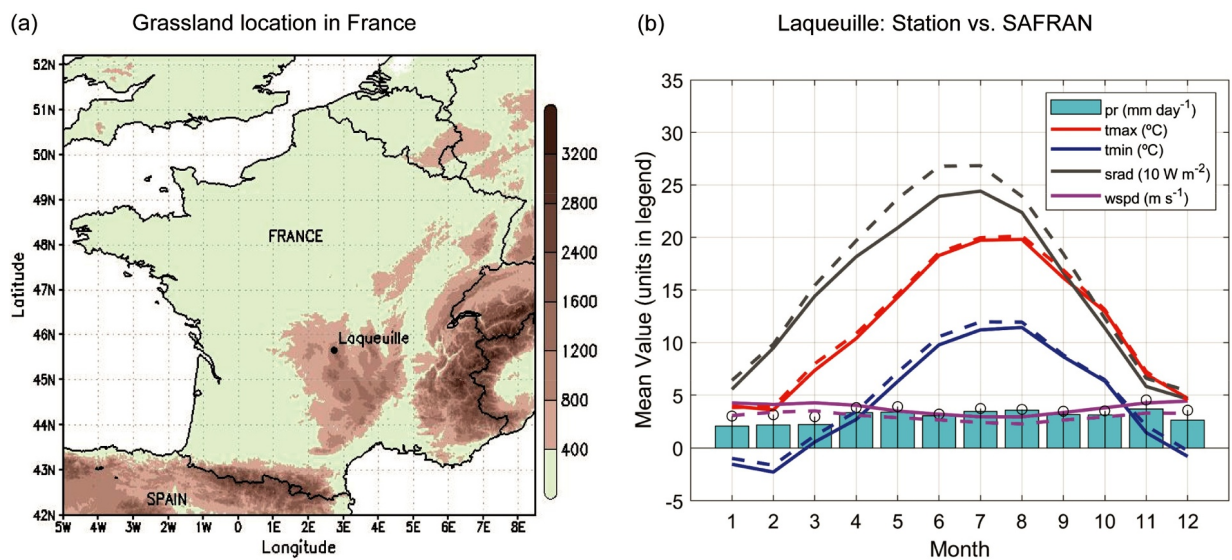

**Fig. 1.** (a) Location of INRA long-term observational grassland site at Laqueuille (Puy-de-Dôme) and its terrain elevation (in m). Source: NOAA NGDC GLOBE: Gridded 1-km. (b) Mean monthly values of surface solar incident radiation (dark gray – 10 W $m^{-2}$), maximum and minimum daily temperatures (red/blue - °C), daily precipitation (cyan - mm day$^{-1}$) and wind speed (magenta – m s$^{-1}$) from the in situ meteorological station (solid lines and bars) and SAFRAN closest grid cell data to Laqueuille (dashed lines and stems). Period 1996–2015 except for solar radiation (2008–2015).

Table 1
Daily mean values of meteorological variables from the INRA in situ station and SAFRAN closest grid cell to Laqueuille. For clarity, the total annual values of precipitation are incorporated. Standard deviation values of all variables are also provided. All values are obtained from the 1996–2015 period except surface solar radiation (2008–2015). Missing values from station data represent at most 0.006% of total for the periods and variables selected.

| Surface weather measurements at Laqueuille | INRA station | | | SAFRAN | | |
| --- | --- | --- | --- | --- | --- | --- |
| | Mean | SD (days) | SD (years) | Mean | SD (days) | SD (years) |
| Solar radiation (W m$^{-2}$) | 148.27 | 96.61 | 6.48 | 163.11 | 102.35 | 7.39 |
| Temperature (°C) | 8.03 | 6.88 | 0.75 | 8.40 | 6.77 | 0.64 |
| Daily precipitation (mm day$^{-1}$) | 2.99 | 6.42 | 0.49 | 3.55 | 6.44 | 0.35 |
| *Annual precipitation (mm yr$^{-1}$)* | *1092* | – | *178* | *1295* | – | *127* |
| Relative humidity (%) | 77.88 | 18.63 | 2.66 | 78.60 | 14.25 | 2.17 |
| Wind speed (m s$^{-1}$) | 3.76 | 1.80 | 0.31 | 2.81 | 1.30 | 0.24 |

## 3. Methods

### 3.1. Statistical methods and teleconnection indices

Climate anomalies were calculated by subtracting the mean value from a timeseries of weather data and then dividing the results by its standard deviation (standardization process). For seasonal/monthly anomalies, the mean of the corresponding season/month was thus subtracted to remove the impact of the seasonal cycle.

To separate contributions from interannual and lower frequency variability (e.g., multi-decadal), a zero-phase spectral 10th-order Butterworth filter with 11-year cut-off period was utilized in high-pass and low-pass modes. The choice of 11-year cut-off period was based on previous studies with a similar focus on interannual and lower-frequency climate variability (Gómara et al., 2016). The Butterworth method (Rabiner and Gold, 1975) has the advantage that it efficiently removes trends in high-pass mode and minimizes undesired edge effects compared to analogous spectral filters (e.g., Lanczos; cf. Supplementary Fig. S1b for an example of surface radiation timeseries).

The methodologies applied to calculate the teleconnection indices are specified in Table S2. A regression of interannual mslp anomalies on the NAO index is provided in Fig. 2a (NAO+ pattern - shadings). SST patterns associated with the Niño3.4, TNA, MED and AMO indices are provided in Fig. 2b (positive phases). Please note that they appear plotted all together in Fig. 2b just for space saving and not to indicate any potential coincidence in time.

Student-t tests that account for the autocorrelation of the timeseries through the calculation of effective degrees of freedom were used for hypothesis testing (Bretherton et al., 1999).

### 3.2. The pasture simulation model (PaSim)

Grassland simulations were performed with the Pasture Simulation model (PaSim). PaSim is a deterministic, biogeochemical grassland-plot model (initially developed by Riedo et al., 1998) incorporating climate data, soil properties, vegetation characteristics, livestock (dairy/beef cattle and sheep) and management, and operating at a point in space on an hourly time step. PaSim simulates multi-year growth of perennial species. It considers the vegetation cover as a single-plant community (although a fixed percentage of legumes can be set to simulate symbiotic N fixation). The model is suited to assess the effect of management practices in terms of fertilization, grazing intensity and duration, and cutting frequency.

The model calculates water, C and N pools and fluxes. Photosynthetic-assimilated C is allocated at each time step to four pools (roots, leaves, stems/sheaths and ears). For each organ, the biosynthesis pathway implies a transition on four age classes from newly produced tissue until senescence. Animal milk production, enteric methane emissions, nitrogen emissions from nitrogen input flows (including returns to soil from animal excreta), and ecosystem respiration are outflow fluxes. Accumulated above ground biomass, if not mown or grazed, enters litter reservoirs. The litter is evenly distributed into the whole soil profile, segregated into its structural and substrate components. The soil organic matter also differentiates between active, slow and passive pools with different decomposition rates according to first-order kinetics. N inputs to the soil include atmospheric deposition, biological fixation by legumes and fertilizer addition. Losses of N occur via pathways that include nitrate leaching, ammonia volatilization and gaseous emissions through microbial conversion of ammonium and nitrate (such as nitrification and denitrification). In the ground process scheme, the soil profile is divided into six layers. In each layer, the kinetics for N flows are driven by C/N ratios with the following values; structural: 150; metabolic: 10; active: simulated in the range 3–14; slow: simulated in the range 12–20; passive: simulated in the range 6–8. Soil temperature and moisture are simulated as a function of soil physical properties and plant water use in each layer of soil.

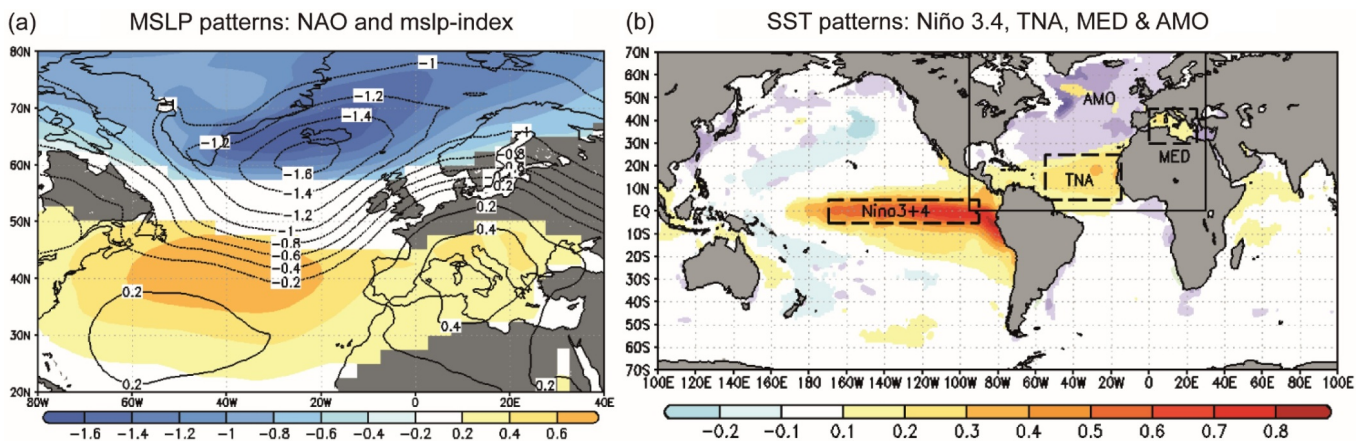

Fig. 2. (a) Regression maps of NAO (shadings: 95% confidence interval; *t*-test) and mspl-index (contours) on interannual mslp anomalies north of 20° N (hPa SD$^{-1}$). (b) Same as (a) but for Niño 3.4, TNA and MED indices on interannual SST anomalies (°C SD$^{-1}$; red/blue shadings) and AMO index on multi-decadal (> 33-yr; magenta/green shadings). The boxes indicate the SST areas considered for the indices. All shadings indicate 99% conf. int. (*t*-test).

The model is enabled for gridded simulations (Vital et al., 2013; Eza et al., 2015) over large-scale areas and for climate change-impact studies (Graux et al., 2011, 2013; Lardy et al., 2013; Ma et al., 2015; Pulina et al., 2018). PaSim is also part of multi-modeling exercises in the frame of international initiatives (Sándor et al., 2015, 2016a,b, 2017, 2018a; Ehrhardt et al., 2018). Sándor et al. (2018b) improved the model to account explicitly for thermal acclimation of plants. In addition to SAFRAN weather data, NOAA Mauna Loa $CO_2$ annual mean data (NOAA-ESRL, 2019) were used as input for PaSim.

### 3.3. Experimental setup: grassland management and model calibration

Alternative grassland management options for modeling were designed with different levels of complexity. Specific details of all performed simulations are provided in Table 2 and Fig. 3.

All runs simulated potential forage yield, whose definition closely mirrors the one used for arable crops, i.e. "the yield of a cultivar when grown in environments to which it is adapted, with nutrients and water non-limiting and with pests, diseases, weeds, lodging, and other stresses effectively controlled" (Evans and Fischer, 1999). In this way, the gap between observed and simulated yield is sensitive to the optimal management conditions (e.g. fertilization, irrigation, cuts), among other potential factors (e.g. model errors).

The choice to use purely simulated forage data rather than empirical data was due to: (i) the short time period covered by the available observational data (~15 years), which hampered inferring any robust conclusion in the context of this work, as generally periods of at least 30 years are desirable to establish robust links and analyze causality between year-to-year climatic and non-climatic variables (Wilks, 2006; IPCC, 2014); and (ii) observed data were influenced by other factors like pests, technical development or research priorities, which obscured the purely climatic origin of the vegetation response that this study aimed to identify. Simulated data has helped to overcome these issues in previous studies, as the most convenient to reveal such interactions (Capa-Morocho et al., 2014, 2016a,b).

The choice of principally simulating mown grasslands in an environment where half of the land is grazed (Rapey, 2016; Estel et al., 2018) instead of using actual empirical grazing data also responded to the need to generate purely climate-dependent timeseries of forage production. For this purpose, the simplest management conditions (cf. MSB in Table 2 and Fig. 3a, which does not include any cutting event) were first evaluated.

Subsequently, management complexity was increased via more frequent cutting events (e.g., 2C, 3C; Fig. 3b and c). Finally, we considered the most complex scenario, denoted 'optimal management' or OM (Fig. 3d), which mimics mowing grassland management generally used by farmers in Europe (Vuichard et al., 2007b). Mown grasslands are not uncommon in the French Massif Central area, where conserved forage (silage and hay) is often utilized by dairy farms to ensure cattle feeding requirements in winter (e.g. Violleau, 1998; Baumont et al., 2011).

Regarding the grassland model setup, a regional 'European' calibration, obtained by calibrating the model against multiple locations (either grazed and/or mown) in Europe (Ma et al., 2015), was applied for the site of Laqueuille. The fact that Laqueuille is a grazed system and not mown as in our simulation, enforced the use of regionally-adjusted parameters instead of site-specific calibrated ones, thus reducing simulation accuracy in a way that could be acceptable to ensure a general applicability of the modeling framework. By accepting the extrapolation principle from Ma et al. (2015), we held it valuable to extrapolate towards a mown grassland in Laqueuille, where rich site-specific soil, plant and weather data were also available to back the study. The choice of Laqueuille for this pilot study was also appealing because this site is conveniently located in the main grassland region of France, the Massif Central.

**Table 2**
Summary of PaSim simulations, including information on N fertilization applications, irrigation, number of cuts per year, cutting or maximum shoot biomass dates (day of year), mean potential forage yield and associated standard deviation (in t DM ha$^{-1}$). Weather input in all simulations is SAFRAN for the period 1959–2015.

| Simulation | Acronym | Description | Mean Yield | SD Yield |
|---|---|---|---|---|
| Maximum Shoot Biomass | MSB | Two N applications triggered on days 14 (17 January) and 120 (30 April) with 160 kg N ha$^{-1}$ for each application (higher N amounts provide same results). Irrigation disabled (as in the rest of simulations if not stated otherwise). Automatic cuts disabled. Although no forage harvests take place in this simulation, the yield is assumed to be equal to the maximum shoot biomass attained in the middle of the growing season. As an example, Fig. 3a provides the simulated potential shoot biomass of year 1982 (maximum around day 175 - 24 June). | 6.87 | 0.9 |
| 2 Cuts | 2C | Two annual forage cuts. The first when the value of maximum shoot biomass is reached (same as in MSB, with varying cutting dates between years). The second when a second maximum after regrowth is attained. Potential forage yield value of each year is retained (Fig. 3b - green line). Additional to the N applications on days 14 and 120 (160 kg N ha$^{-1}$ each), a third N application is triggered after the first annual cut. | 8.45 | 1.4 |
| 3 Cuts | 3C | Three annual forage cuts triggered. Dates of cutting events are fixed for all years on days 150 (30 May), 225 (13 August) and 300 (27 October) to try better capture seasonal effects in forage growth (e.g., summer hydrological stress). Potential forage yield value is retained. Fertilization events take place on days 14, 120 and after the two first forage cuts (Fig. 3c – year 1982). | 8.51 | 2.1 |
| 3 Cuts + Irrigation | 3CI | Same as 3C with optimum irrigation conditions. The optimal irrigation method relies on a water stress index from PaSim. | 8.68 | 2.2 |
| Optimal Management | OM | The optimal management routine from Vuichard et al. (2007b) is utilized to simulate the potential forage yield under the possible closest circumstances to reality. Cutting dates vary each year and the annual potential forage yield values are retained. The first cut of each year is triggered right before plant senescence. The subsequent cuts take place whenever plant growth rate declines during 10 consecutive days and there is enough available forage to collect (> 1 t DM ha$^{-1}$) after at least 30 days of regrowth. Fertilization events are optimized based on a N nutrition index (N application amounts reduced to minimum non-limiting forage growth levels). In Fig. 3d, the management for year 1982 is also provided as an example. | 7.60 | 1.5 |

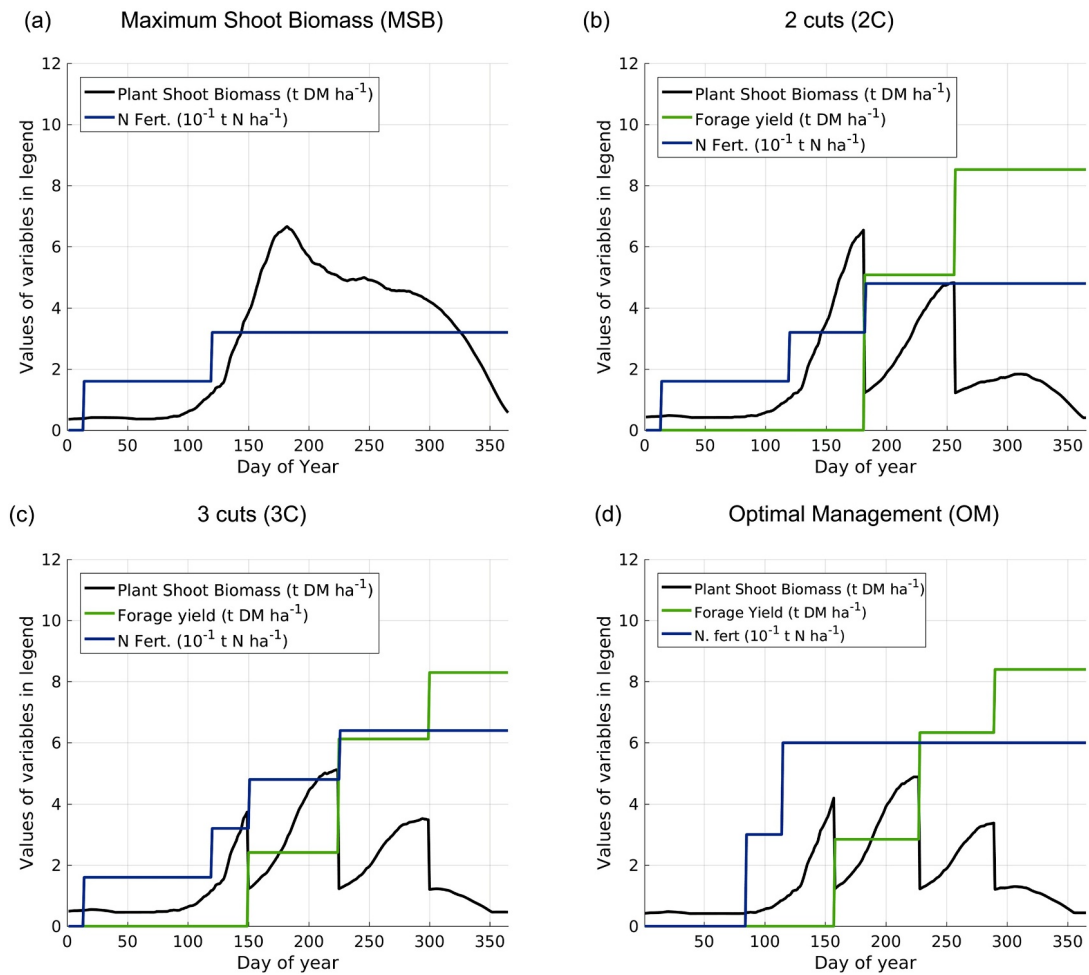

Fig. 3. (a) Grassland management in MSB simulation (example of year 1982). Black solid line: Potential plant shoot biomass (t DM ha$^{-1}$). Blue solid line: Cumulative fertilization year-round ($10^{-1}$ t N ha$^{-1}$). (b) Same as (a) but for 2C management. Green solid line: potential forage yield accumulation (t DM ha$^{-1}$). (c) Same as (b) but for 3C management. (d) Same as (b) but for optimal management.

### 3.4. Statistical seasonal forecast model of potential forage yield

A statistical seasonal forecast model of potential forage yield at Laqueuille was developed. A stepwise approximation was followed for adding or removing terms in the multi-linear regression model. The method automatically searches for sets of oceanic/atmospheric predictive terms improving forecast skill through forward selection and background elimination. At each step, the p-value of an F-statistic is utilized to select 'in' and 'out' predictors (Draper and Smith, 1998). Next, a leave-one-out cross-validation method was applied, where the model was trained from predictor/predictand data combinations from all years (1959–2015) except the year being predicted at each training stage (Allen, 1971).

### 4. Results and discussion

This section first illustrates the temporal evolution of grassland productivity under alternative management options and analyzes possible correlations between forage production and biophysical factors (weather variables and $CO_2$ concentration; Section 4.1). Then, the contributions of trends/low-frequency (Section 4.2) and high-frequency (Section 4.3) climate variability on forage yield are presented separately. Results and discussion of the developed statistical seasonal forecast model are provided in Section 4.4.

### 4.1. Temporal evolution of grassland productivity

With higher number of cuts (with or without irrigation; 3CI/3C), simulations present higher mean productivity than with limited cuts (2C) or maximum shoot biomass annual values (MSB; Fig. 4a and Table 2).

However, the optimal management simulation (OM) does not generate the highest potential forage yield, owing to internal biophysical constraints on the simulated forage production (e.g., water soil bearing capacity may not adequately support loads, or cutting events after 30 days of regrowth are only triggered whenever sufficient forage yield is available for harvesting, > 1 t DM ha$^{-1}$).

The standard deviation values (Table 2) appear to arise from different temporally-dependent contributions (Fig. 4a). The first and most evident contribution is associated with trends and/or potential low frequency oscillations in forage yield. For instance, MSB increases by 29% (+1.74 t DM ha$^{-1}$) between 1959–1979 and 1995–2015. The second contribution is due to interannual yield changes, which is more evident under more intensive management options (3C, OM).

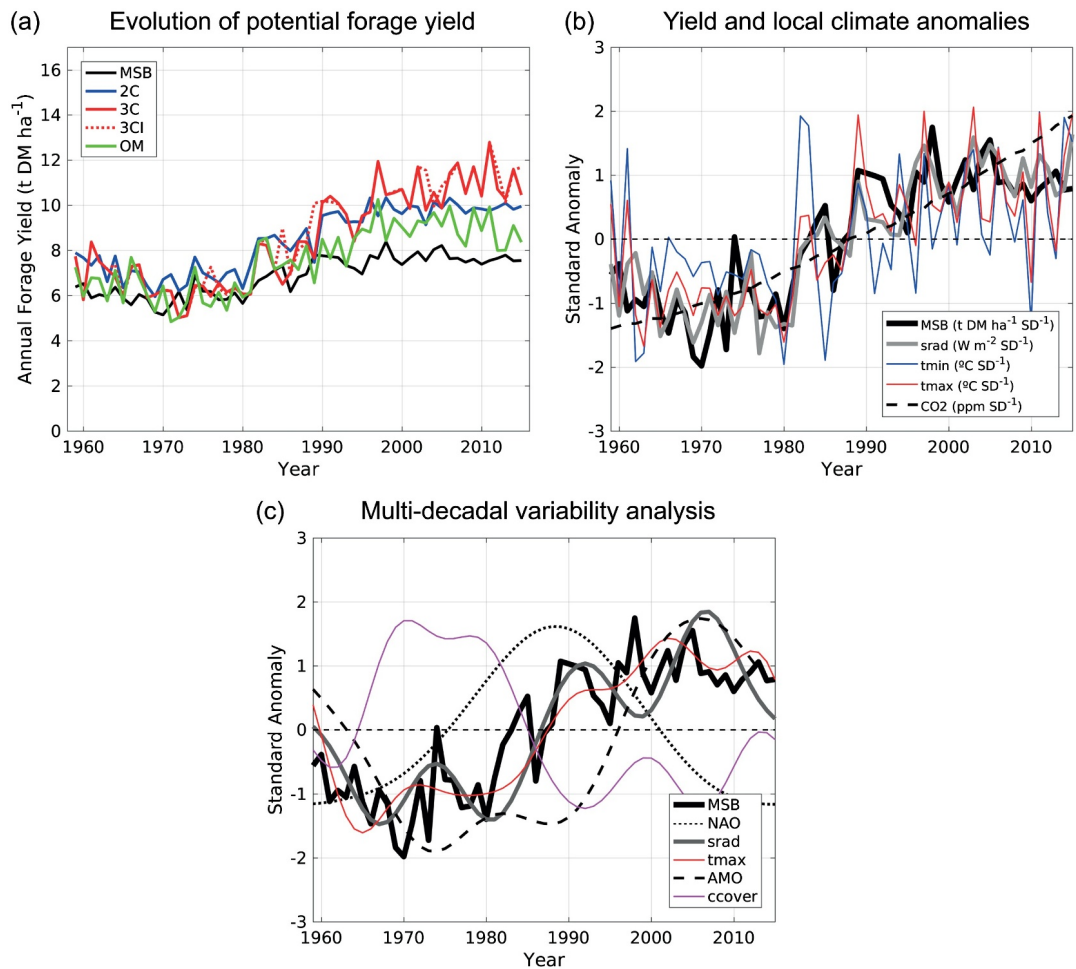

**Fig. 4.** (a) Time evolution of annual potential maximum shoot biomass and forage yield (2C, 3C, 3CI and OM) for PaSim simulations (t DM ha$^{-1}$). Period 1959–2015. (b) Time evolution of annual averaged standardized anomalies of surface radiation (srad; W m$^{-2}$ SD$^{-1}$), Tmax/Tmin (°C SD$^{-1}$), precipitation (Pr; mm day$^{-1}$ SD$^{-1}$), $CO_2$ concentration (ppm SD$^{-1}$) and potential maximum shoot biomass (MSB simulation; in t DM ha$^{-1}$ SD$^{-1}$). (c) Same as (b) but for low-pass filtered (> 11-yr) anomalies of SAFRAN srad, Tmax and NCEP cloud cover (ccover; in % SD$^{-1}$) averaged over mainland France. The multi-decadal NAO and AMO indices (> 33-yr) appear also overlaid.

To analyze possible climate-related causes for these changes, Fig. 4b provides the yearly averaged standardized anomalies of observed climatic conditions at the grassland together with yield for the MSB simulation, which is the simplest management scheme. Anomalies of surface radiation (srad), maximum and minimum temperatures (Tmax/Tmin) and $CO_2$ concentration appear to covary with yield, both in terms of year-to-year and longer-term changes.

The correlation values between these climatic variables (plus precipitation) and yield are provided in Table 3 (left values), srad being the main driver of forage yield variability in all simulations. For completeness, correlation coefficients were also calculated between yield and other climatic variables also utilized by PaSim to estimate water balance (water vapor pressure and wind speed), but they did not provide anything worth commenting on.

These results are consistent with the Alpine-like climate of the grassland site, where incident solar radiation is limited due to frequent rainfall (cf. Table 1 and Fig. 1b) and temperatures are low due to the high altitude (1040 m a.s.l.). Higher values of $CO_2$, srad and Tmax/Tmin (all highly correlated with yield) are known to stimulate plant photosynthesis and promote shoot biomass growth in Alpine environments (Menzi et al., 1991; Ainsworth and Rogers, 2007). Logically, applying frequent forage cuts during the year (3C, OM) introduces higher variability in simulated grassland productivity than just measuring the maximum shoot biomass attained in the middle of the growing season of each year (standard deviation values in Table 2), which typically occurs in June (MSB; cf. Fig. 3a). This becomes especially evident after the decade of the 1990s, when atmospheric and climatic conditions considerably improved (higher $CO_2$, Tmax/Tmin and srad; Fig. 4b), raising the yield ceiling and making strategic management a much more important driver of productivity (Fig. 4a).

**Table 3**
Correlation values between annual averaged anomalies of potential maximum shoot biomass / forage yield and SAFRAN climatic conditions at the grassland (period 1959–2015): incident solar radiation (srad), maximum/minimum daily temperature (Tmax/Tmin), precipitation (Pr) and $CO_2$ concentration. Values to the left/center/right are obtained from raw/low-pass/high-pass filtered timeseries. Boldface indicates 95% confidence interval (t-test).

| Correlation | Srad | Tmax | Tmin | Pr | $CO_2$ |
| --- | --- | --- | --- | --- | --- |
| **MSB** | **0.87 / 0.96** / 0.26 | **0.72 / 0.92** / 0.06 | **0.39 / 0.75** / 0.01 | **−0.31 / −0.45** / −0.25 | **0.82 / 0.88** / 0.01 |
| **2C** | **0.89 / 0.98** / 0.12 | **0.71 / 0.93** / −0.06 | **0.44 / 0.80** / 0.10 | −0.19 / **−0.39** / 0.01 | **0.86 / 0.90** / −0.01 |
| **3C** | **0.87 / 0.95 / 0.31** | **0.79 / 0.90 / 0.47** | **0.56 / 0.79 / 0.54** | −0.20 / **−0.32** / −0.13 | **0.86 / 0.91** / −0.09 |
| **3CI** | **0.93 / 0.97 / 0.68** | **0.86 / 0.93 / 0.73** | **0.57 / 0.82 / 0.56** | **−0.35 / −0.42 / −0.52** | **0.87 / 0.93** / −0.09 |
| **OM** | **0.84 / 0.96** / 0.23 | **0.77 / 0.89 / 0.46** | **0.59 / 0.79 / 0.55** | −0.14 / **−0.28** / 0.01 | **0.78 / 0.86** / −0.03 |

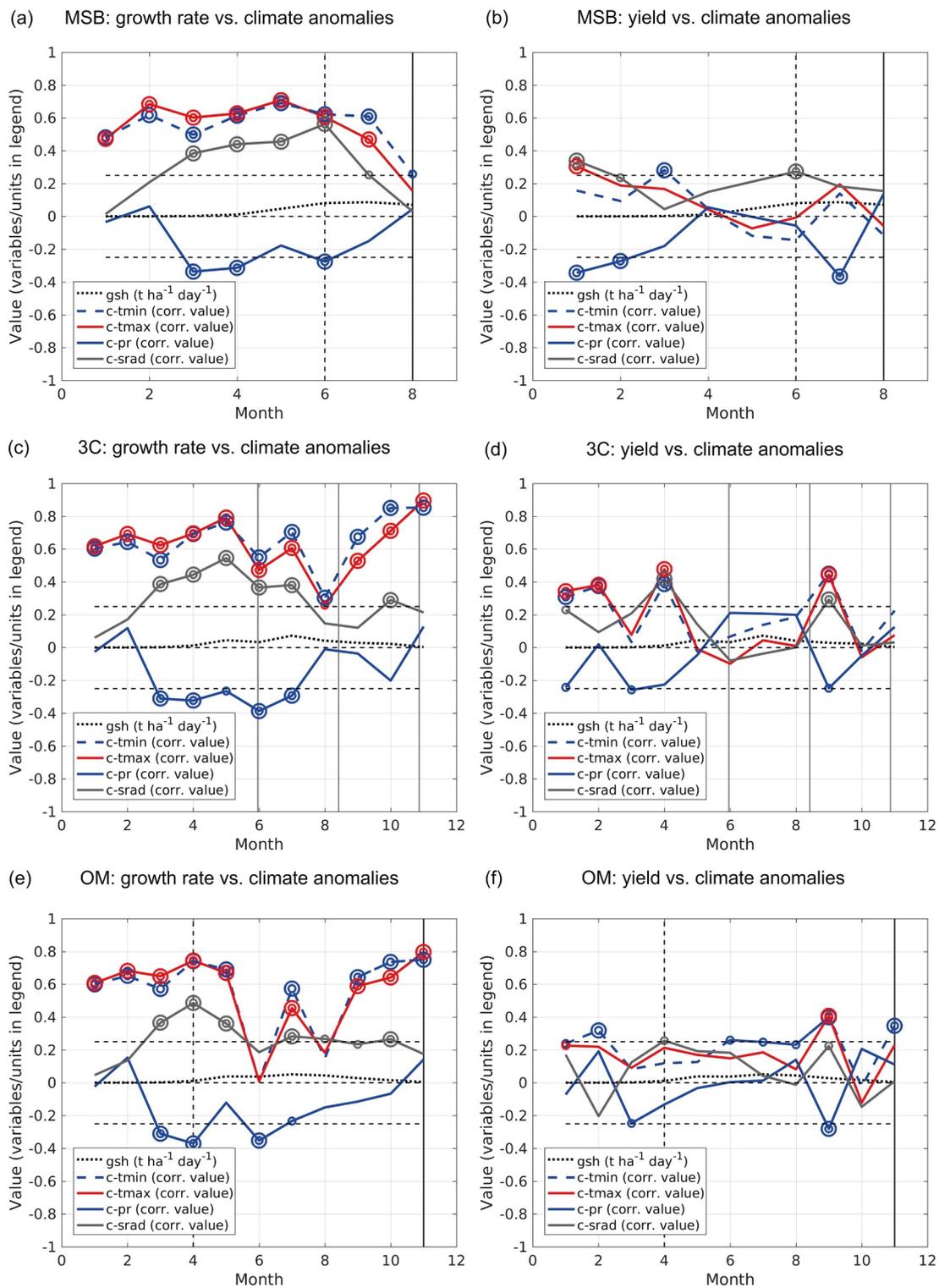

**Fig. 5.** (a) Correlation values between high-pass filtered monthly anomalies (12 values per year) of shoot growth rate (gsh; t ha$^{-1}$ day$^{-1}$) and meteorological variables at the grassland site: srad (W m$^{-2}$), Tmax/Tmin (°C) and Pr (mm day$^{-1}$). 90/95% confidence intervals (*t*-test) marked with small/big circles. Horizontal lines provide an estimate of the range of non-statistically significant correlations. Vertical lines indicate the starting (dashed) and end (solid) period of maximum shoot biomass measurements (MSB simulation). The mean growth rate values for each month (gsh; t ha$^{-1}$ day$^{-1}$) are provided in the dotted black line. (b) Same as (a) but for correlations between annual averaged potential yield (MSB; one value per year) and monthly climatic anomalies (12 values per year). (c)–(d) Same as (a)–(b) but for the 3C simulation. Vertical solid lines denote fixed cutting events in time. (e)–(f) Same as (a)–(b) but for the OM simulation. The vertical lines indicate the starting (dashed) and end (solid) period of automatic OM cutting events.

For precipitation, the correlation with yield is, in general, slightly negative and not statistically significant ($p > 0.05$). Again, this is consistent with the high rainfall and low radiation at the site. First, rainfall abundance prevents relevant water stress on grassland growth.

Second, yield increases when radiation and surface temperature increase, which in turn are usually associated to lower cloudiness and precipitation. Both interactions together explain the precipitation-yield response.

### 4.2. Impact of trends and low-frequency climate variability on grassland productivity

The correlations between low-pass filtered timeseries of yield and annual climatic anomalies are provided in Table 3 (central values). Positive and statistically significant ($p < 0.05$) correlations can be observed between yield and srad, Tmax/Tmin and $CO_2$, and negative for Pr. This is particularly evident between srad and the different biomass simulations (0.95–0.98 correlation values). Thus, the long-term changes of the variables explaining a higher fraction of yield variability, srad, Tmax and $CO_2$, were subsequently analyzed.

In Fig. 4c, the raw timeseries of MSB yield is provided together with low-pass filtered srad and Tmax. Apart from the filtering, the main difference between Fig. 4c and b is that srad and Tmax timeseries (in gray/red) are based on spatial averages over mainland France [5° W-8° E, 42° N-52° N] and not just Laqueuille. This indicates that the increase in srad and Tmax between 1959–1979 and 1995–2015 (Fig. 4c) was associated to a large-scale feature.

Regarding the potential causes for the large-scale increase in srad over France, several authors have pointed to the global 'dimming and brightening' effect related to anthropogenic emission of aerosols (Wild et al., 2005; Sánchez-Lorenzo et al., 2009). However, the level of detail of this study does not allow disclosing the radiative effect of aerosols (absorption or scattering). Instead, multi-decadal changes in cloud cover fraction seem to be the main driver of srad changes (cf. Fig. 4c - gray and magenta lines). The multi-decadal NAO index is overlaid in Fig. 4c because NAO is one of the main drivers of multi-decadal cloudiness variability over the North Atlantic (Trigo et al., 2002), due to its influence on low pressure systems' trajectories (Ulbrich and Christoph, 1999; Gómara et al., 2014, 2016). The index may depict a remarkable variability (Wang et al., 2012; Pinto and Raible, 2012), but the observed behavior is not consistent with srad changes. Another teleconnection pattern with significant influence on European climate at these timescales is the AMO (Fig. 2b; Knight et al., 2005; Wollings et al., 2012). The AMO index (also overlaid in Fig. 4c) does appear to covary with srad and cloud cover changes. Nevertheless, it is still only a brief snapshot of a long-term cycle. A negative AMO-cloud cover relation in the North Atlantic, mediated by SST meridional gradients and storm track activity, has been recently pointed out by Vaideanu et al. (2017).

The positive trends in Tmax/Tmin and $CO_2$ concentrations (Fig. 4b and c) are in line with global warming trends (IPCC, 2014). Trends of Tmax values show a similar increase at Laqueuille and over mainland France when disaggregated among seasons (Figs. S1c-d). The same results were obtained for Tmin (data not shown).

Investigating further these aspects, though interesting, would deviate too much from the initial objectives of this study, and could be arranged later (e.g., through climate-grassland sensitivity model experiments) as a natural evolution of what is presented here.

### 4.3. Impact of interannual and seasonal climate variability on grassland productivity

#### 4.3.1. Relations between grassland shoot growth rate, forage yield and climate anomalies

To analyze the influence of interannual climate variability on grassland productivity, all considered timeseries in this section were high-pass filtered (if not stated otherwise). The correlation values between climate and forage yield anomalies are shown for all simulations in Table 3 (right values). At interannual timescales, the main drivers of forage productivity are again srad (MSB/3C/3CI simulations) and Tmax/Tmin (3C/3CI/OM). Despite still being statistically significant, the observed correlations are clearly lower at interannual timescales than for the raw timeseries. In addition, correlation coefficients vary among simulations (e.g., srad/Tmax/Tmin values for MSB and 3C/3CI; Table 3).

To identify the seasons of the year in which climate conditions present a stronger influence on forage growth, Fig. 5 (left column) provides the correlation values between monthly climate and shoot growth rate anomalies (MSB, 3C and OM simulations). Results for 2C and 3CI are also available in supplementary Fig. S2. As expected, Tmax/Tmin and srad positive anomalies at the grassland are robustly linked ($p < 0.05$) with enhanced forage growth rates for long periods of the year. For clarity, the monthly mean shoot growth rates appear overlaid in Fig. 5 (dotted black line in panels). Again, the negative correlation with precipitation is a consequence of the high and homogeneous rainfall, and the positive effect of sunnier and warmer weather conditions on forage growth at this mountainous location. In addition, the periods in which these relations are stronger are sensitive to the timing of cutting events and seasonal changes in climate conditions. For MSB (Fig. 5a), the period of the strongest climate-growth rate links spans from January to July. This is because maximum shoot biomass values tend to occur between June and August. For 3C (Fig. 5c), as the three annual cuts span from June to November, the influence of Tmax/Tmin and srad on growth rate is extended until the end of the year (from January to November).

During the summer months this influence appears slightly reduced though. This behavior is also observed in the 2C simulation (Fig. S2a) and might be consistent with increased water demand during summer, thus requiring a more balanced equilibrium between Tmax/Tmin/srad and Pr effects. This hypothesis is confirmed in Fig. S2c, where the 3CI simulation (which performs optimal irrigation removing Pr dependency) shows higher correlation values with srad and T during the summer months compared to 3C (Fig. 5c). For OM, results are very similar to the 3C simulation (Fig. 5c and e).

Next, the anomalies of total annual forage yield (1 value per year) were correlated against monthly climatic anomalies (12 values per year). Therefore, for each simulation, the same yield timeseries was correlated against the different monthly climatic anomalies (Fig. 5, right column). In general, the links between forage yield and climate anomalies are notably weaker compared to forage growth rates. This is because correlation values are no longer based on concurrent monthly anomalies. Instead, yield values provide combined information of climate conditions over long periods of the year and the efficiency of grassland management to materialize increased forage growth rates into actual yield. Thus, simulations indicate that the most intensively managed grasslands are better linking srad/Tmax/Tmin and yield anomalies, especially during spring and fall. This can be clearly observed when comparing 3C/OM/3CI simulations (Figs. 5d/f/S2d) with MSB/2C ones (Figs. 5b/S2b). During the summer months similar results point to increased Pr dependency, caused by increased water demand habitual of this season (e.g. 3C; Fig. 5d). This increased Pr dependency is logically more evident for correlations based on longer temporal windows (compare blue line in Fig. 5c -monthly- versus d - annual).

#### 4.3.2. Relations between forage yield and large-scale monthly climate anomalies

To confirm whether the climate conditions fostering grassland productivity were specific to the site location (e.g., absence of upslope fogs) or related to any large-scale atmospheric circulation pattern, the correlation between forage yield at Laqueuille and SAFRAN climate anomalies was assessed over mainland France (Figs. 6 and 7; left and central columns). The choice of variables (Tmax/Tmin/srad), simulations (3C/OM) and months (January/February/April/September) responds to Fig. 5 outcomes: 3C and OM simulations provide the strongest links between meteorological surface variables and yield during the selected months. Besides, 3C and OM simulations are respectively the most productive and closest to real-life ones (Table 2). As it can be observed, statistically significant correlations (in stippling) extend over vast areas. This implies that warmer temperatures and enhanced srad that produce higher forage yields at Laqueuille are generally present at the same time over large areas of mainland France. This is particularly more evident for 3C, as the regions of significant positive correlations are larger than for OM, for the latter being, in some cases, even of opposite sign, but not statistically significant (Fig. 7e). For a better view, all correlation maps with SAFRAN Tmax/Tmin/srad from January to December are provided in Figs. S3 (3C) and S4 (OM) zoomed over France.

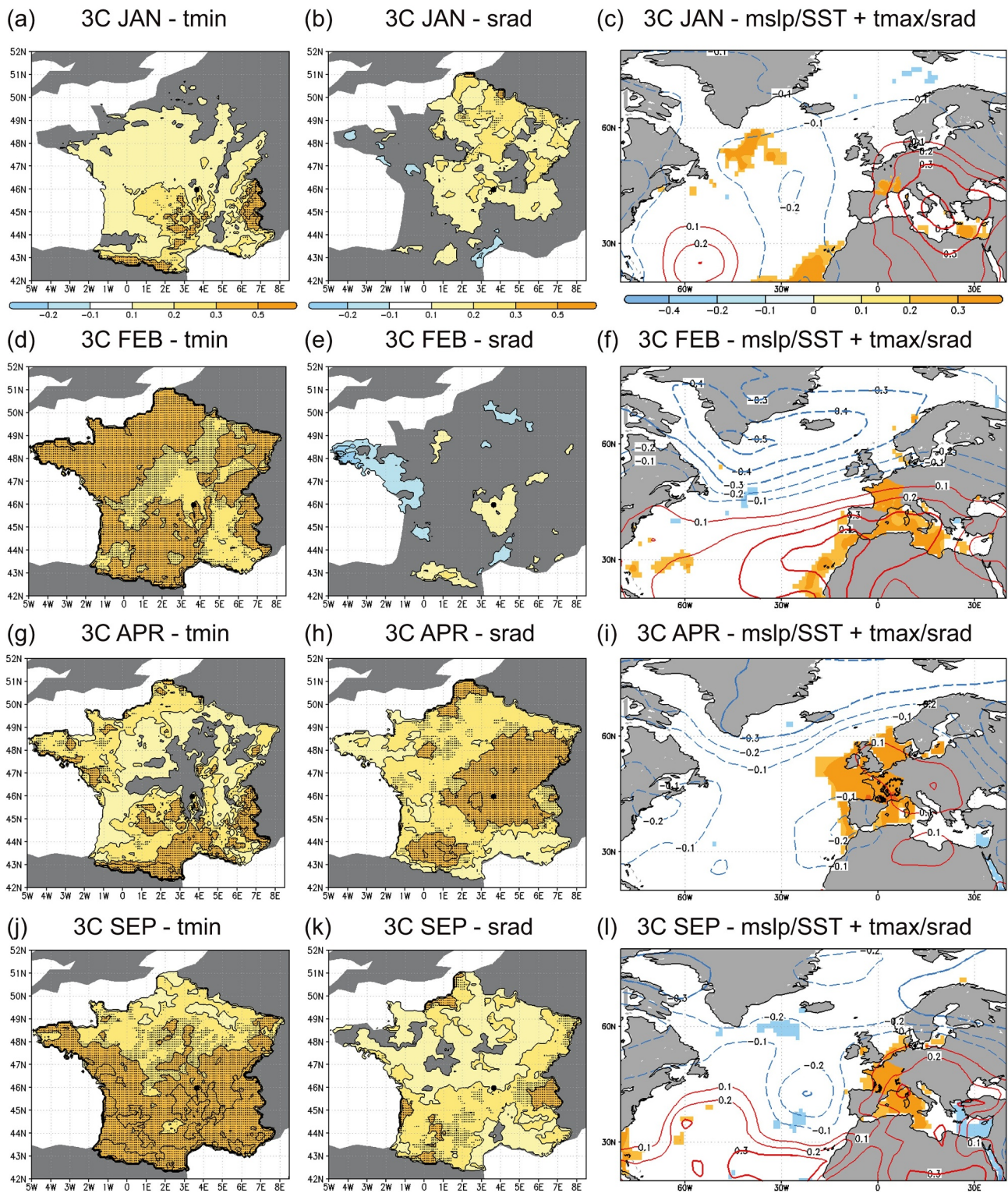

**Fig. 6.** Correlation values between Laqueuille high-pass filtered potential forage yield (3C simulation) timeseries and January monthly averaged anomalies of: (a) SAFRAN Tmin (shadings and contours; °C). 95% conf. int. in stippling. (b) SAFRAN srad (W m$^{-2}$). (c) NCEP mslp (contours; hPa), HadSST SSTs (shadings over the seas; °C), SAFRAN Tmax (shadings over France) and SAFRAN srad (black contours over France - same as in (b)). 95% conf. int. in all shadings, red/blue thick contours and all black contours. (d)–(l) Same as (a)–(c) but for February, April and September. Period 1959–2015.

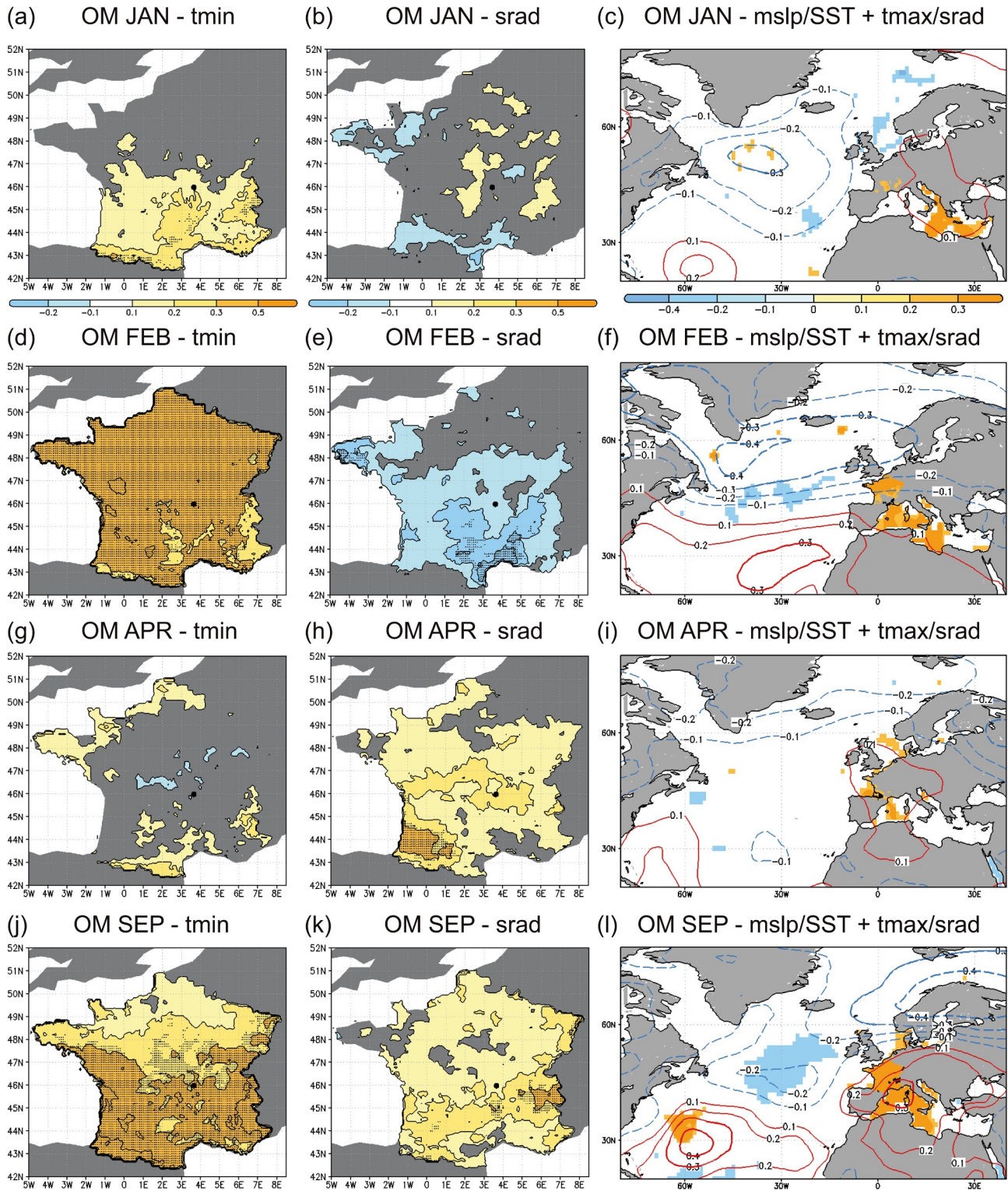

Fig. 7. Same as Fig. 6 but for the OM simulation.

To analyze the origin of these large-scale patterns, Figs. 6 and 7 also provide analogous correlation maps of yield versus mslp (NCEP; contours) and SST (HadSST; shadings) anomalies over the North Atlantic (right columns). Years of enhanced yield are generally associated with positive mslp anomalies over the Mediterranean and sub-tropical North Atlantic, and with negative mslp anomalies near Greenland and Scandinavia. These mslp correlation maps notably resemble the positive phase of the NAO, but with centers of action slightly displaced southeastward. To test this relationship, we constructed an index of the observed mslp structure as the spatial correlation between the composite of mslp maps from Figs. 6 and 7 (eight members in total; pattern in Fig. 2a - contours) and mslp high-pass filtered monthly anomalies for the period 1959–2015. Hereafter, we refer to it as the mslp-index (cf. also Table S2).

The correlation value between NAO and mspl-index is 0.81 ($p < 0.01$), while the correlation with the East Atlantic pattern, a similar mode to NAO but with centers of action shifted south, is 0.25 ($p < 0.1$) (Barnston and Livezey, 1987; their Fig. 9d). These results indicate that the identified mslp pattern appears to enhance Tmax/Tmin and srad over large areas of France, including the Massif Central, which in turn promotes forage yield at Laqueuille. This is consistent with positive mslp (anticyclonic) anomalies over the area, which are typically associated with more stable weather; sunnier and warmer conditions during daytime. For completeness, the same maps covering the North Atlantic area are provided for all months for 3C and OM simulations in Figs. S5 and S6, respectively.

Regarding SSTs, Figs. 6 and 7 also depict a positive robust connection between enhanced 3C/OM yield and a warming of the marine waters near France (Mediterranean, North Sea and far eastern North Atlantic). This is especially evident in the months of February and September (Figs. 6 and 7f, l). As described in the introduction, these SST conditions are typically linked to higher temperatures over Europe (Feudale and Shukla, 2007; García-Serrano et al., 2013).

To test the sensitivity of results to the choice of atmospheric re-analysis (NCEP, ERA-40, ERA-Interim; cf. Table S1) and provide a global view on the relation between mslp/SST and 3C/OM yield anomalies, in Figs. S7–S8 the analogous correlation maps are provided. In all cases, the 3C/OM yield timeseries are the same (cf. Table 2 - forced by SAFRAN) and the only factors that vary are the mslp anomalies from each re-analysis database and the period utilized to compute the correlations (cf. Table S1). For simplicity, in Fig. 8 the main outcomes from Figs. S7–S8 panels are outlined.

A first conclusion is that correlation patterns of mslp are quite similar among re-analyses. A second conclusion is that additional SST patterns to MED also appear to influence simulated forage yield, but only for the 3C simulation. This is the case between 3C yield and TNA SSTs in January (Fig. 8a–c) and with pacific equatorial SSTs (ENSO) in May (Fig. 8g–i). More precisely, supplementary Figs. S7 and S9 reveal that ENSO-3C yield connection takes place between April and July (Fig. S7j–u) and that TNA influences yield from September of the previous year (hereafter y-1; Fig. S9i–l) to January of harvest year (y0; Fig. S7a). Additionally, these SST-yield relationships appear to evolve in time. For instance, the influence of ENSO on 3C yield is much stronger in ERA-Interim (period 1979–2015) than in NCEP (period 1959–2015; compare Fig. 8g and i).

The differences observed between 3C (Fig. S7) and OM (Fig. S8) simulations may be caused by the much more complex management and limitations (soil bearing capacity, sufficient yield for harvesting) in the OM simulation. The fact that 3C cuts are fixed in time every year and OM ones not could also contribute to sharpen ENSO/TNA SST climatic impacts on yield. For instance, if ENSO influenced climate conditions during summer, this signal could be captured by the 2nd cut in 3C (always covers the period June to August). For OM, as cuts are variable in time, this relation could get attenuated.

As a summary of the relations identified so far on NCEP (the longest re-analysis), in Fig. 9 the lagged correlation values of yield versus mslp (NAO, mslp-index) and SST (TNA, ENSO, MED) teleconnection indices are provided for 3C (Fig. 9a) and OM (Fig. 9b). As expected, the mspl-index is significantly correlated with 3C yield on February, April and September y0 ($p < 0.05$). Similar results are obtained for NAO and MED on the same months. Niño3.4 is significantly correlated with yield in May and June y0 and TNA from September y-1 to January y0. For OM, only the links between yield and mspl-index/MED remain for February and September.

### 4.3.3. Influence of teleconnection patterns on forage yield variability

To assess the individual influence of the identified teleconnections on the observed climatic and yield anomalies at Laqueuille (Figs. 5–8) and propose a bio-physical mechanism for forage growth, a reverse methodology was applied.

For this purpose, the indices of the teleconnection patterns were directly correlated with monthly anomalies of climate variables at Laqueuille during the previous and following months over France (SAFRAN) and the whole North Atlantic (NCEP). The selected teleconnections for analysis, based on Fig. 9a,b results, were: (i) mspl-index in February and September (3C and OM); (ii) MED in February and September (3C and OM); (iii) ENSO in May-June (3C); and (iv) TNA from September y-1 to January y0 (3C).

Results for (i) are provided in Fig. 9c,d. In Fig. 9c correlations between February mspl-index and January to June climate anomalies indicate that this teleconnection pattern is associated with increased Tmin/Tmax and srad over Laqueuille from January to May. In the same figure the correlations of Feb/mspl-index with the NAO and itself for the previous and following months (black solid/dotted lines) indicate that the persistence of this atmospheric pattern is of just 2 months (February and March).

For completeness, the correlation maps between Feb/mspl-index and concurrent monthly anomalies of mslp (NCEP; red/blue contours), SST (HadSST; shadings), moisture flux at 850 hPa (NCEP; arrows), Tmax (SAFRAN; shadings over France) and srad (SAFRAN; black contours over France) are shown in Fig. 9d. For the calculation of moisture flux, the zonal and meridional wind anomalies at 850 hPa were multiplied by the same pressure level specific humidity values (in $m\ g\ s^{-1}\ kg^{-1}$). As expected, the anticyclonic conditions associated with mspl-index over central France promote higher temperatures and srad (Fig. 9d) and seem responsible for the enhanced forage growth rates at Laqueuille between February and May (Fig. 9c - green line).

In Fig. 9e,f the analogous results are provided for the February MED index. In this case, the autocorrelation of the Feb/MED SST index for the previous and following months indicates a higher temporal persistence of the anomalies in time (magenta line in Fig. 9e). This is due to the high thermal capacity of SSTs (Bjerknes, 1969). A warmer Mediterranean is also associated with increased temperatures and precipitation at Laqueuille in January and February and higher forage shoot growth rates from January to April (Fig. 9e).

The associated large-scale correlation maps with Feb/MED (Fig. 9f) indicate that a warmer Mediterranean seems to be associated with enhanced temperatures over central Europe and a dipolar mslp structure over the North Atlantic. This mslp structure resembles the East Atlantic pattern and is suggested to be a potential driver of interannual SST winter anomalies in the Mediterranean (Skliris et al., 2011; Zveryaev and Hannachi, 2012).

Attending to the timing and persistence of mspl-index and MED in Fig. 9c and e, results suggest that the MED SST anomalies grow earlier (January) but are not necessarily responsible of the mspl-index pattern onset in February (black line in Fig. 9e). However, whenever mspl-index is established in February, the associated anticyclonic conditions over the Mediterranean could as well contribute to the persistence of the positive SSTs over the area (positive feedback – magenta line in Fig. 9c). The combined effect of mspl-index and MED patterns seems responsible for the enhanced forage growth rate values in Fig. 9c until May (green line), a behavior which cannot be explained by the effect of the mspl-index alone.

Regarding precipitation, the enhanced values in January/February associated to a warmer Mediterranean (Fig. 9e) could be explained by the increased moisture flux values over France in Fig. 9f (arrows). The positive SST anomalies also present near the European Atlantic coast may enhance surface evaporation and humidity, which is subsequently transported towards France by the anomalous wind flow.

For completeness, the analogous of Fig. 9c–f but for the month of September are provided in Fig. S10 (June to November period). The results for September are highly similar to those of February and indicate that mspl-index and MED teleconnection patterns are as well responsible of increased shoot biomass productivity at Laqueuille in August and September.

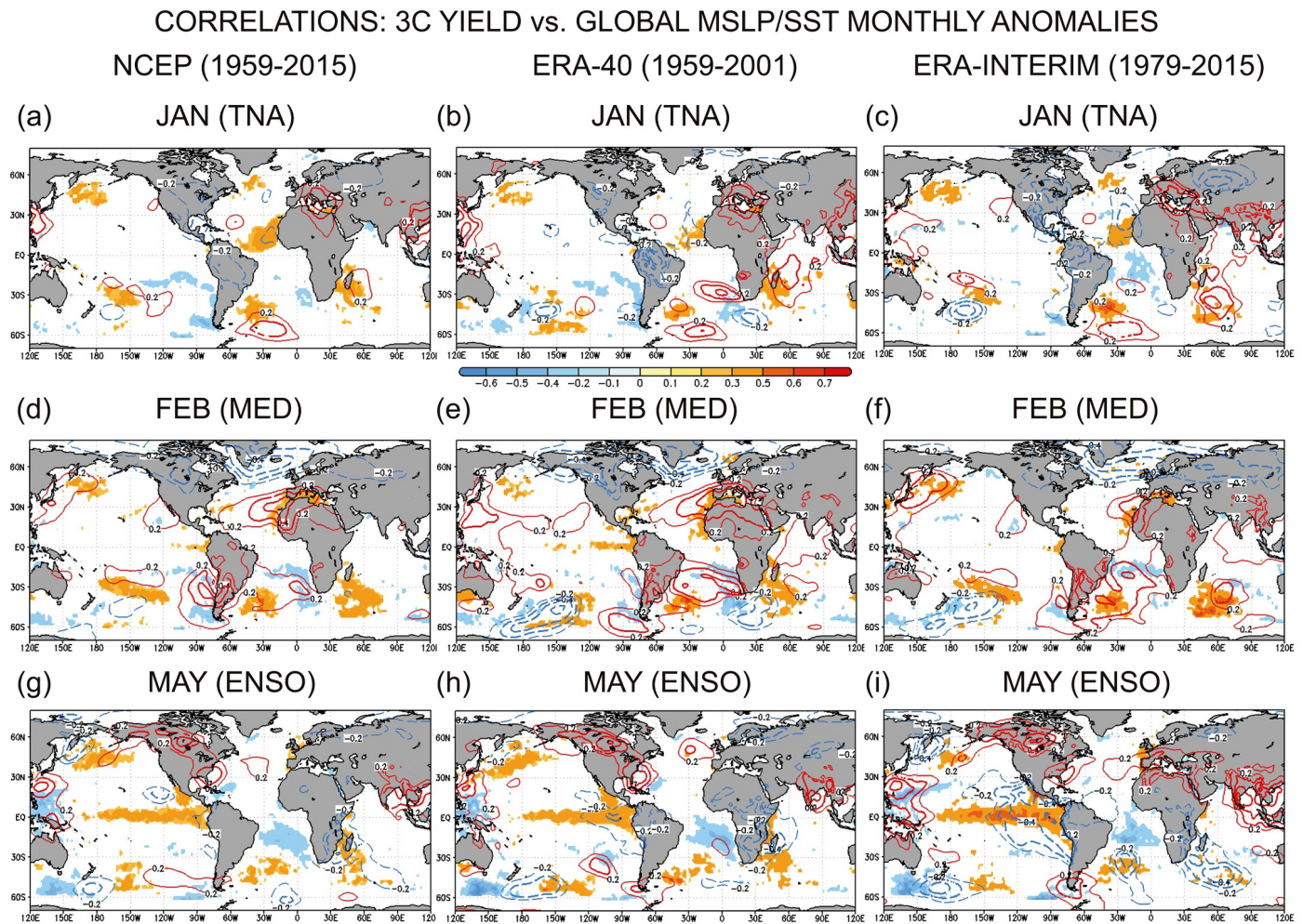

**Fig. 8.** Correlation maps between high-pass filtered 3C potential forage yield (1 data point per year) and January monthly averaged anomalies of SST (HadSST; 95% conf. int. in shadings) and mslp (95% conf. int. in thick contours) from (a) 1959–2015 NCEP, (b) 1959–2001 ERA-40 and (c) 1979–2015 ERA-Interim. (d)–(f) Same as (a)–(c) but for February. (g)-(i) Same as (a)-(c) but for May.

The lead/lag correlations for ENSO and TNA indices did not provide clear and persistent statistically significant anomalies of Tmin, Tmax, srad or shoot growth rates during the harvest year on NCEP (not shown). Therefore, a direct joint climate and eco-physiological mechanism for forage growth was hard to establish. Typically, a positive TNA in summer and autumn is associated with NAO- during the next winter and enhanced precipitation over southwestern Europe (Okumura et al., 2001; Rodríguez-Fonseca et al., 2006). In this line, one potential mechanism for enhanced yield may be associated with increased soil moisture storage during the previous autumn and winter to forage collection, thus ensuring water supply within a certain range for plant growth in the following months (Craine et al., 2012). For ENSO, as the highest correlation values were found in May and June (Fig. 9a), one possible mechanism could be that the associated signal influenced the precipitation regime in Laqueuille during summer and fall (Shaman, 2014; his Fig. 1c). In this line, several studies have shown that increased precipitation over southern/central Europe occurs under El Niño conditions (Mariotti et al., 2002; Park, 2004; Shaman, 2014), enhancing crop productivity over the Iberian Peninsula (Capa-Morocho et al., 2014; 2016a,b,c). Consequently, water stress conditions during the summer months (detected in Fig. 5c,d) may be minimized and forage growth could be promoted.

Regarding these mechanisms, a much deeper analysis may be needed to establish the physical links between the SST precursors and climate anomalies at Laqueuille. Instead, in the next section an effort was made to forecast 3C and OM grassland annual forage yield values at Laqueuille based on climatic predictors identified so far.

### 4.4. Statistical seasonal forecast of forage yield

#### 4.4.1. Model performance

A statistical multi-regression model was built to predict annual values of yield anomalies for the 3C and OM simulations. The purpose of the model is to forecast, for each year, whether the forage productivity at the grassland will be 'good' (above average) or 'bad' (below average) several months ahead. To this aim, local climatic monthly anomalies and large-scale monthly teleconnection indices from y0 and y-1 were considered as predictors. As in the previous sections, all considered timeseries were high-pass filtered. The considered predictors for the model were: (i) local monthly anomalies at Laqueuille: Tmax, Tmin, srad and Pr; and (ii) large-scale monthly teleconnection indices: NAO, mspl-index, MED, TNA and Niño3.4. The initial choice of predictors was based on Figs. 5–8 outcomes. Next, a stepwise approximation was followed for adding or removing terms in the multi-linear regression model (Draper and Smith, 1998).

The construction of a prediction model was first carried out for the 3C simulation. As annual cuts of forage are always triggered on days 150 (30 May), 225 (13 August) and 300 (27 October), atmospheric and oceanic predictors were only considered until the month of May y0. Therefore, the forecast windows with cuts are 0 (cut1), 2.5 (cut2) and 5 (cut3) months, with the total annual yield value always measured after cut3. In this sense, the contributions of the three cuts into the total yield annual value are, respectively, 36, 43 and 21% (1959–2015 averages). These numbers indicate a high degree of actual prediction (64% of total forage; sum of 2nd and 3rd cuts) under the current model configuration. Based on the selected variables, the stepwise method returned 6 predictors providing model skill: [TNA September y-1, mslp-index January y0, mslp-index February y0, MED February y0, Tmax Laqueuille April y0 and Niño3.4 May y0]. Next, a leave-one-out cross-validation method was applied.

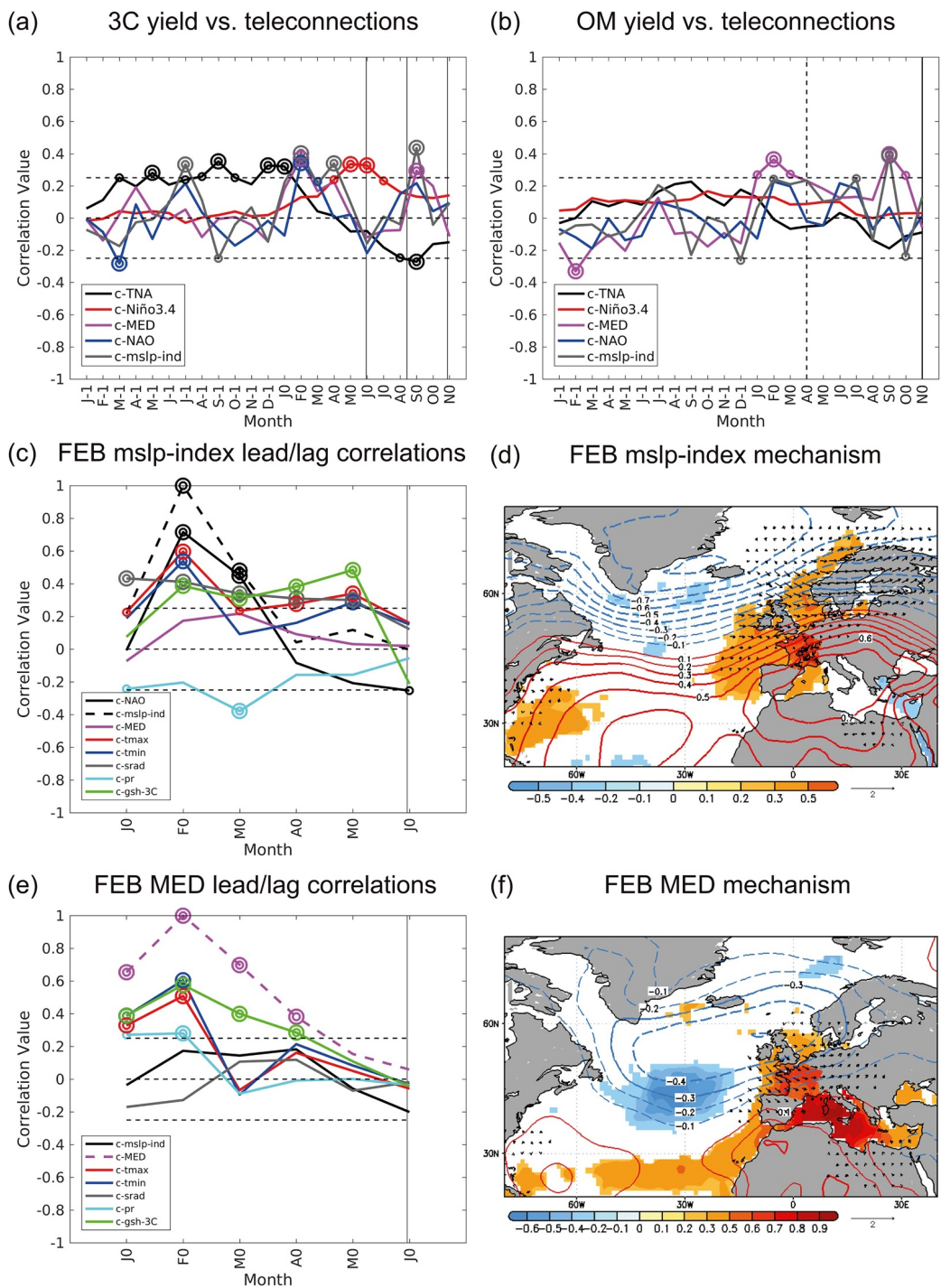

Fig. 9. (a) Same as Fig. 5d but for correlations between TNA, Niño3.4, MED, NAO and mslp-index monthly indices and interannual forage 3C yield anomalies. Considered months: January year −1 to December year 0 (harvest year). (b) Same as (a) but for OM yield anomalies. (c) Same as (a) but for lead/lag correlations of February mslp-index with itself, NAO, MED and Tmin/Tmax/srad/Pr and 3C shoot growth rate (gsh) monthly anomalies at Laqueuille from January to June year 0. (d) Same as Fig 6c but for correlations with February mslp-index instead of 3C annual yield. Arrows indicate correlations above 0.1 for positive moisture advection at 850 hPa (NCEP; m g s$^{-1}$ kg$^{-1}$). (e)–(f) Same as (c)–(d) but for February MED index.

Fig. 10a provides the 'observed' (simulated by PaSim) and predicted timeseries of forage yield anomalies based on the generated statistical model. The model performs quite well, providing a skill in correlation of 0.55 ($p < 0.01$, t-test), a root mean square error percentage (% RMSE) of 14.6% and a 71.43% hit rate in determining whether the yield of a given year will be above or below average. Please note that for the latter the baseline prediction skill is 50%. In the form of a tercile-forecast (33% superior, central or inferior values), the model returns a 48.21% hit rate, the baseline being 33%.

The same analysis was subsequently performed for the OM simulation. In this case, the number of selected predictors by the stepwise method was notably lower: [MED February y0 and Tmin Laqueuille July y0]. This was somewhat expected due to the low number of statistically significant correlation values between OM yield, climatic anomalies (Fig. 5f) and teleconnection indices (Fig. 9b). To improve the balance between the actual forecast window and model performance, the period for potential predictors was extended until July y0. For the OM simulation, cutting dates are variable among years, triggered from April to November. That provides a 0 to 4-month forecast window between atmospheric/oceanic predictors and yield, but also some temporal overlapping between April and July. Indeed, the model could have been improved by introducing yield values from the first cuts of the year as predictors, but this would have gone in essence against the initial objectives of this paper. The results for the regression model after the leave-one-out cross-validation are provided in Fig. 10b. In this case, the skill in correlation is 0.41 ($p < 0.01$), %RMSE 17.6% and the hit rates of above/below average annual harvest and tercile-forecasts are 67.86% and 44.64%, respectively (50% and 33% baselines).

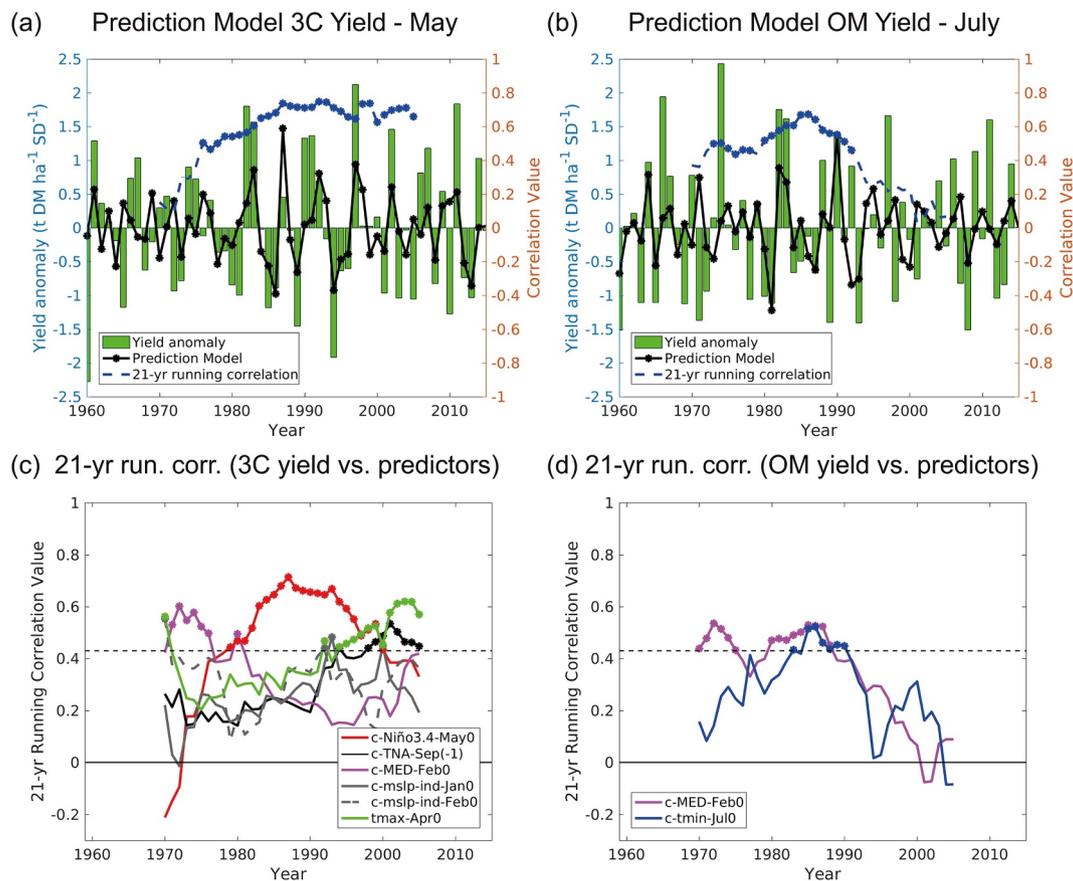

**Fig. 10.** (a) Stepwise regression model with leave-one-out cross-validation between potential 3C yield anomalies (predictand: bars in t DM ha$^{-1}$ SD$^{-1}$) and selected predictors (black line) until May of harvest year [TNA September y-1, mspl-index January y0, mspl-index February y0, MED February y0, Tmax Laqueuille April y0 and Niño3.4 May y0]. (b) Same as (a) but for OM. Predictors are selected until July of harvest year [MED February y0 and Tmin Laqueuille July y0]. The 21-yr centered running correlations between predictand/predicted timeseries appear also overlaid (blue lines: 95% confidence interval with filled circles). (c) 21-yr centered running correlations between identified climatic model predictors and 3C yield. Filled circles indicate 95% conf. int. (*t*-test). (d) Same as (c) but for OM simulation.

*4.4.2. Analysis of stationarity in model predictability*

This section analyzes the evolution in time of the links between forage yield and climate predictors. This was motivated by the results obtained in Fig. 8g–i, where the link between ENSO-3C yield appeared to be sensitive to the period considered. For this purpose, the 21-year centered running correlations between the 'observed' and 'predicted' 3C/OM yield timeseries are also overlaid in Fig. 10a,b (blue line; $p < 0.05$ in circles).

For 3C (Fig. 10a), a much higher skill in correlation was found between 1980 and 2015 (max. 0.75) compared to 1959–1980 (min. 0.1). To analyze the causes underlying this behavior, in Fig. 10c the same running correlations were applied between 3C yield and model predictors separately. Results indicate that the main source of 3C yield predictability seems to be ENSO and that this link is not stationary in time. During the periods in which the link between ENSO-3C yield is weaker, other sources provide effective forecast skill (MED 1970–1980; TNA/Tmax 1990–2005).

For OM, a non-stationary relationship between climate predictors and simulated yield is also seen (Fig. 10b), with much higher correlation values over the 1980s and 1990s (up to 0.67) than in the first period of the 21st century (down to 0.03). These changes can be attributed to the time evolving links between MED/Tmin and OM yield (Fig. 10d).

The results outlined in this sub-section could be associated with multi-decadal changes in the shape and location of teleconnections (Jung et al., 2003; Gómara et al., 2016). For instance, the impact of ENSO on European climate has been shown to be non-stationary in time during the 20th century (Mariotti et al., 2002; Greatbach et al., 2004; Rodríguez-Fonseca et al., 2016), as is also the case of ENSO impact on Iberian Peninsula crops (Capa-Morocho et al., 2016a,b). An extended assessment considering temporal windows of enhanced prediction opportunity and their underlying factors may require a more complete analysis at this point, which is beyond the scope of this paper. Therefore, it is left as an interesting path for future research.

*4.4.3. Final remarks on seasonal forage yield predictions*

The results presented here provide support to a future implementation of an in-season forage productivity prediction system at French and European levels, which could be of interest to governments, administrations, chambers of commerce and insurance companies (JPI-Climate, 2019). Grassland production systems are complex and the relationship between forage production and the associated environmental factors are difficult to examine. However, biogeochemical grassland models are ideal for simulating forage production due to their ability to represent the non-linear response of the modelled output to changes of soil, climate and management conditions. To our knowledge, this is the first modeling study to explore the connections between large-scale oceanic–atmospheric circulation and local forage production in Europe. By mobilizing a suite of simulations, it has provided an improved understanding of the response of grassland biomass production to purely climatic patterns. With focus on the French Massif Central, the modeling approach used in this study suggests that: (1) seasonal prediction of grassland productivity from climatic predictors is possible several months prior to cutting; and (2) the links between climate predictors and grassland yield are also not stationary in time.

The relationship between site-specific forage production and NAO and SST patterns identified in this study warrants continue further monitoring to substantiate the role of atmospheric and oceanic changes on grassland dynamics beyond management changes. Further evidence supporting this relationship would permit reasonable inferences about forage production in the form of seasonal (1–5 month) forecasts.

While suggesting the possibility of using grassland models in support of Climate Services, the present study also shows limitations. Our findings are based purely on simulated data to avoid the noise from other factors (e.g., pests, technical development); therefore, they lack evaluation against measured outputs. This is consistent with previous studies that used observed data only for model calibration and validation (Capa-Morocho et al., 2014; 2016a,b), as simulated production data are considered as the most appropriate to avoid masking plant-weather interactions by non-related factors. Moreover, the model used in this study is only a possible realization of alternative approaches to grassland modeling. However, we think that the 'European' calibration/validation version of PaSim applied for the study site give us, at a minimum, confidence on the reliability of forage production estimates (Ma et al., 2015). Another limitation is that grazing schemes were not assessed; this first analysis could only simulate mowing, even though grazing is common in the area. The methodological simplifications adopted in the research (e.g., simulation of potential forage yield, mowing practices) were to some extent necessary owing to the complex nature of grassland systems, as not a single harvest event had to be forecasted, but a continuous biomass production, which in turn, may follow different paths depending on management. Our methodology enables progress towards operative forecasts of forage production. Since this is the first study of this type, it provides ground for further developments, and we expect that a great deal more research will be necessary to unravel the complex relationships at the interface between climate, vegetation and management at different timescales.

## 5. Conclusions

The potential links between climate variability, atmospheric and oceanic teleconnection patterns and forage production of a mown permanent grassland system in the French Massif Central (Laqueuille) were analyzed in this study. Following a biogeochemical grassland modeling approach, evidence was provided that skillful seasonal forecasts of forage productivity at the site are achievable using oceanic and atmospheric predictors (e.g., Sea Surface Temperature indices).

This paper provides essential elements to implement a timely and accurate grassland forecasting system in France and, after substantiation in other grassland systems (either with mowing, grazing or both), through gridded simulations over Europe. Such a tool, together with monitoring services (e.g., Copernicus Sentinel satellites; ESA, 2019) and farm economic models (Britz et al., 2014), would facilitate to increase on-farm seasonal managerial flexibility, mitigate risks and adopt cost-effective local and macro-economic plans (cf. Chen et al., 2017). For instance, by proactively adapting cutting or grazing schedules, stocking rates, irrigation and fertilization plans or production and fallow schemes.

**Declaration of Competing Interest**

None.

**Acknowledgments**

We thank the National Centers for Environmental Prediction, the European Center for Medium-Range Weather Forecasts and Météo-France/Hymex/MISTRALS for providing the NCEP, ERA-40, ERA-Interim and SAFRAN re-analyses. We thank the UK Met-Office Hadley Center for the HadSST database. This study was supported by the project MACSUR - Modeling European Agriculture with Climate Change for food Security (FACCE-JPI), funded by Instituto Nacional de Investigación y Tecnología Agraria y Alimentaria (INIA), the meta-program ACCAF (Adaptation of agriculture and forests to climate change) of the French National Institute for Agricultural Research (INRA) and the project PRE4CAST (CGL2017-86415-R) of the Spanish Ministry of Economy and Competitiveness (MINECO). We also thank the Laqueuille site as part of the SOERE-ACBB project (http://www.soere-acbb.com), funded by the French National Infrastructure (https://www.anaee-france.fr). Iñigo Gómara was supported by MINECO (Juan de la Cierva-Formación contract; FJCI-2015-23874) and Universidad Politécnica de Madrid (*Programa Propio – Retención de Talento Doctor*). Iñigo Gómara's research stay at INRA UREP (September-December 2017) was also funded by UPM (Programa Propio - Ayudas al personal docente e investigador para estancias breves en el extranjero). We would like to thank Olivier Darsonville, Rémi Perrone, Iris Lochon, Katja Klumpp, Catherine Picon-Cochard (INRA UREP), Pere Quintana-Seguí (Observatori de l'Ebre) and Roberto Suárez-Moreno (Columbia University) for their helpful cooperation during the progress of this study. Finally, we are indebted to the two anonymous reviewers, whose pertinent comments and suggestions have contributed to substantially improve this manuscript.

**Supplementary material**

Supplementary material associated with this article can be found, in the online version, at doi:10.1016/j.agrformet.2019.107768.